\documentclass{JAC2003}
 
\usepackage[dvipdfm]{graphicx}
\usepackage{booktabs}
\usepackage{amsmath}
\usepackage{amssymb}

\begin{document}
\title{Unruh radiation and Interference effect
\thanks{ Based on a poster presentation 
by Y.Yamamoto and  \cite{IYZ}.}}
\author{Satoshi Iso\thanks{satoshi.iso@kek.jp}, Yasuhiro Yamamoto\thanks{yamayasu@post.kek.jp} and Sen Zhang\thanks{zhangsen@post.kek.jp}, KEK, Tsukuba, Japan}

\maketitle

\begin{abstract}
A uniformly accelerated charged particle feels the vacuum as thermally
excited and fluctuates around the classical trajectory. 
Then we may expect additional radiation besides the 
 Larmor radiation.
It is called Unruh radiation.
In this report, we review the calculation of 
the Unruh radiation with an
emphasis on the interference effect
between the vacuum fluctuation and the radiation from the 
fluctuating motion.
Our calculation is based on a stochastic treatment of the particle
under a uniform acceleration.
The basics of the stochastic equation 
are reviewed in another report 
in the same proceeding \cite{IYZ2}.
In this report, we mainly discuss the radiation and the interference
effect. 
\end{abstract}
\section{STOCHASTIC ALD EQUATION} 
The Unruh radiation is the additional radiation expected to be emanated
by a uniformly accelerated charged particle \cite{ChenTajima}.
A uniformly accelerated observer 
feels the quantum vacuum as thermally excited with the Unruh temperature
$T_U=\hbar a/2\pi ck_B$. Hence as the ordinary Unruh-de Wit detector,
a charged particle interacting with the radiation field can be 
expected to fluctuate around the classical trajectory.
Is there additional radiation  associated with this
fluctuating motion? It is the issue of the present report.

In order to formulate the dynamics of such fluctuating motion,
we make use of the stochastic technique.
Namely, we solve a set of equations of 
the accelerated particle and the radiation field
in a semiclassical approximation.
By semiclassical, we mean that the radiation field is treated
as a quantum field while the particle is treated classically.

Since the accelerated particle dissipates its 
energy through the Larmor radiation, the equation of motion 
contains the radiation damping term.
This is the Abraham-Lorentz-Dirac (ALD) equation.
Furthermore, since the accelerated particle feels the 
Minkowski vacuum as thermally excited, a noise term
is also induced in the equation of motion.
The stochastic equation of the accelerated charged 
particle  is called the stochastic ALD equation
and derived by \cite{JohnsonHu}.

We consider the scalar QED whose action is given by
\begin{align}
 S[z,\phi, h] =& -m \int d\tau \  \sqrt{\dot{z}^{\mu} \dot{z}_\mu}
+ \int d^4x \ \frac{1}{2}(\partial_\mu \phi)^2 \nonumber \\
&  + \int d^4x \ j(x;z) \phi(x) .
\end{align}
where 
\begin{align}
 j(x;z) = e \int d\tau \ \sqrt{\dot{z}^\mu \dot{z}_\mu} \delta^4 (x-z(\tau)),
\end{align}
We choose the parametrization $\tau$ to satisfy $\dot{z}^2=1$.

By solving the Heisenberg equation for $\phi$, 
 we get the  stochastic ALD equation for the charged particle:
\begin{align}
 m \dot{v}^\mu 
 - F^\mu - \frac{e^2}{12\pi} (v^\mu \dot{v}^2 + \ddot{v}^\mu) 
  =
 - e \overrightarrow{\omega}^\mu \phi_h(z)
\label{stochastic-particle}
\end{align}
where $v^\mu = \dot{z}^\mu$. 
The dissipative term corresponds to loss of energy through the radiation
and it is called the radiation damping term.
On the other hand, the noise term comes from the Unruh effect, namely,
interaction of a uniformly accelerated particle with the thermal bath 
of the radiation field.

We can easily solve the dynamics of small fluctuations of 
the transverse velocities 
$v^i  = v^i_0 + \delta v^i$  
in terms of the quantum fluctuations of the field  $\phi_h$ 
(or its Fourier tranformed field $\varphi$) as
\begin{align}
 \delta \tilde{v}^i(\omega) = e h(\omega) \partial_i \varphi(\omega),
 \label{solution}
\end{align}
where 
\begin{align}
 \delta v^i(\tau) &= \int \frac{d \omega}{2\pi} 
 \delta \tilde{v}^i(\omega) e^{-i \omega \tau}, \\ 
 \partial_i \phi_h(\tau) &= 
   \int \frac{d \omega}{2\pi} \partial_i \varphi(\omega) 
	     e^{-i \omega \tau} \\
h(\omega) &= \frac{1}{-im \omega +\frac{e^2}{12\pi}(\omega^2 +a^2)}.	     
\label{varphi}
\end{align}
In the following, as an ideal case we consider a uniformly accelerated
charged particle in the scalar QED, 
and investigate the radiation from such a particle.
The main issue is the effect of interference.

\section{RADIATION and INTERFERENCE}
Now we calculate the radiation emanated from 
the uniformly accelerated charged particle.
First let's consider the 2-point function of the radiation field.
Since the field is written as a sum of the 
quantum fluctuation (a homogeneous solution)  $\phi_h$ and the 
inhomogeneous solution in the presence of the 
charged particle $\phi_I$, the 2-point function is given by
\begin{align}
& \langle \phi(x) \phi(y) \rangle - \langle \phi_h(x) \phi_h(y) \rangle \\ &= 
   \langle \phi_{I}(x) \phi_h(y) \rangle
  +\langle \phi_h(x) \phi_{I}(y) \rangle 
  +\langle \phi_{I}(x) \phi_{I}(y) \rangle
\nonumber .
\end{align}
The Unruh radiation estimated in \cite{ChenTajima} is
contained in  $\langle \phi_{I} \phi_{I} \rangle$,
which include the Larmor radiation. 
We need special care of the interference terms.
As discussed  in \cite{Sciama},
the interference terms  
$\langle \phi_{I} \phi_h \rangle + \langle \phi_h \phi_{I} \rangle$
may possibly  cancel the Unruh radiation 
in $\langle \phi_{I} \phi_{I} \rangle$ after the thermalization 
occurs. 
The cancellation is explicitly shown for an internal detector, 
but it is not obvious whether the same cancellation occurs 
for the case of a charged particle we are considering. 

The inhomogeneous solution of the scalar field is written as
\begin{align}
 \phi_{I} (x) &= 
   e\int d\tau G_R (x -z(\tau)) = \frac{e}{4\pi \rho (x)} .
  \label{trans-inh} \\
 \rho (x) &= 
  \dot{z}(\tau^x_{-}) \cdot (x -z(\tau^x_{-})),
  \label{rhodef}
\end{align}
where $\tau^x_-$ satisfies 
$(x-z(\tau^x_{-}))^2 = 0, \quad x^0 > z^0(\tau^x_{-})$,
which is the proper time of the particle 
whose radiation travels to the space-time point $x$. 
Hence, $z(\tau_-^x)$ lies on an intersection between the particle's
world line and the light cone extending from the observer's position $x$
(See Fig 1).
We write the superscript $x$ to make the $x$ dependence of $\tau$ explicitly.

The particle's trajectory is fluctuating and expressed as
$z=z_0 + \delta z+ \delta^2 z +\cdots$ where we have expanded
the tragectory with respect to the interaction with the
radiation field (i.e. $e$).
Then $\rho$ is also expanded as
$\rho = \rho_0 + \delta \rho + \delta^2 \rho + \cdots$
and (\ref{trans-inh}) becomes
\begin{align}  
 \phi_{I} 
 = \frac{e}{4\pi \rho_0} 
 \left( 1 -\frac{\delta\rho}{\rho_0} 
   + \left( \frac{\delta\rho}{\rho_0} \right)^2 
	- \frac{\delta^2 \rho}{\rho_0} + \cdots 
 \right).
 \label{phiexpand}
\end{align}
The first term is the classical potential, 
but since the particle's trajectory deviates from the classical one, 
the potential also receives corrections.

\subsection{Inhomogeneous part}
By inserting the expansion (\ref{phiexpand}), 
the correlator of the inhomogeneous solution $\phi_I$ becomes 
\begin{align}
& \langle \phi_{I}(x) \phi_{I}(y) \rangle \label{inh-inh-result} \\ &=
  \left(\frac{e}{4\pi} \right)^2 \frac{1}{\rho_0(x)\rho_0(y)} 
\nonumber \\ &\times 
  \biggl( 
    1 + 
	 \frac{\langle \delta \rho(x) \delta \rho(y)\rangle}{\rho_0(x)\rho_0(y)} + 
	 \frac{\langle ( \delta\rho(x) )^2 \rangle}{\rho_0^2(x)} + 
	 \frac{\langle ( \delta\rho(y) )^2 \rangle}{\rho_0^2(y)} 
  \biggr) .
\nonumber
\end{align}
The first term  gives the Larmor radiation.
The other terms correspond to the radiation induced by the  fluctuations
of the particle's motion.

The calculations of these terms are easy, because one can write
$\langle \delta \rho \delta \rho \rangle$ in terms of 
$\langle \delta\dot{z}^i\delta\dot{z}^i \rangle 
=\langle \delta v^i \delta v^i\rangle$, which can be obtained 
by solving the dynamics of fluctuations 
in the stochastic ALD equation \cite{IYZ,IYZ2}. 
They become
\begin{align}
& \langle \phi_{I}(x) \phi_{I}(y) \rangle \label{EqIninScl} \\
  &=
  \left( \frac{e}{4\pi} \right)^2 \frac{1}{\rho_0(x)\rho_0(y)}
\biggl[ 1 +
  e^2 
  \int \frac{d\omega}{2\pi} 
  |h(\omega)|^2 I_S(\omega)
  \nonumber \\ 
  & 
  \times \Bigl( 
     \frac{x^i y^i e^{-i\omega (\tau^x_- -\tau^y_-)}}{\rho_0 (x) \rho_0 (y)}
	 + \frac{x^i x^i}{\rho_0^2 (x)}
	 + \frac{y^i y^i}{\rho_0^2 (y)} \Bigr)  \biggr] \nonumber .
\end{align}
Since we are considering the fluctuating motion whose frequency
is smaller than the acceleration,
$I_S$ can be approximately given by $a^3/12\pi^2$.

\begin{figure}[t] 
\begin{center}
\includegraphics[width=20em]{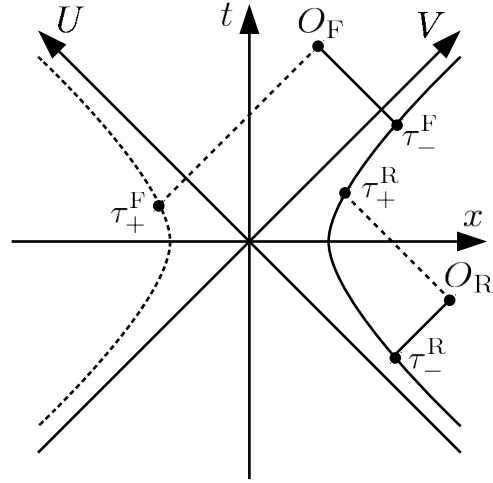}
\caption{
The hyperbolic line in the right wedge denotes the world line of the particle. 
The points $O_{\textrm{F}}$ and $O_{\textrm{R}}$ are observers in the future 
and right wedges, respectively.
For an observer in  the right wedge, 
the light-cone of the observer has two intersections with the world line, 
and the proper time of the intersections are
given by $\tau^R_{\pm}$. 
For an observer in  the future wedge, there is only one intersection 
on the particle's real trajectory which corresponds to $\tau^F_-$.
The other solution $T_+^{\textrm{F}}=\tau_+^{\textrm{F}}+i \pi/a $ is complex. 
One may interpret this complex proper time  as
the intersection between the light-cone of the observer and the world line 
of a virtual particle with a real proper time $\tau_+^{\textrm{F}} $ 
in the left wedge. The superscript letters $R$ or $F$ are used to distinguish
two different observers, but we do not use them in the body of the paper
to leave the space for the observer's position $x$.}
\label{FigRind}
\end{center}
\end{figure} 
\subsection{Interference Term}
The calculation of  the interference terms is a bit more involved.
The inhomogeneous solution $\phi_I$ is expanded as (\ref{phiexpand}).
Since the leading term which has a nonvanishing correlation with the 
quantum fluctuation $\phi_h$ is the second term, we have
\begin{align}
& \langle 
    \phi_{I}(x) \phi_h(y) \rangle + \langle \phi_h(x) \phi_{I}(y) 
  \rangle \nonumber \\
  &= 
  - \frac{e}{4\pi} 
   \left( 
	   \frac{\langle \delta\rho(x) \phi_h(y) \rangle}{\rho_0^2(x)} 
	 + \frac{\langle \phi_h(x) \delta\rho(y) \rangle}{\rho_0^2(y)} 
	\right).
\end{align}
The fluctuation of the distance
$\delta \rho$ is written in terms of $\delta v^i$ which is the solution 
of the stochastic ALD equation (\ref{solution}),
and we obtain 
\begin{align}
  \langle \phi_h(x) \delta \rho(y) \rangle
  &= 
  -e y^i \int \frac{d\omega}{2\pi} e^{-i\omega \tau^y_-} h(\omega) 
  \langle \phi_h(x) \partial_i \varphi(\omega) \rangle \nonumber .
\end{align}
The integrand can be written as
\begin{align}
  \langle \phi_h(x) \partial_i \varphi (\omega) \rangle 
  &=
  \int d\tau e^{i\omega \tau} 
  \left(
    \frac{\partial}{\partial y^i} \langle \phi_h(x) \phi_h(y) \rangle 
  \right)_{y=z(\tau)}  \nonumber \\
&= -\int d\tau e^{i\omega \tau} 
  \left(
    \frac{\partial P(x,\omega)}{\partial x^i} \right),
\end{align}
where 
\begin{align}
  P(x, \omega) =
    \int d\tau \frac{e^{i\omega \tau} }
    {(x^0 -z^0(\tau) -i\epsilon)^2 -(x^1 -z^1(\tau))^2 - x_{\perp}^2 }.
    \nonumber
\end{align}
$x_\perp ^2 = (x^2)^2 +(x^3)^2$ is the transverse distance. 
The $\tau$ integral can be calculated by the contour integral.
The residues are located where the invariant length between 
the observed point $x$ and a point on the particle's trajectory
vanishes.
The condition is nothing but the condition that the radiation field
propagates on the light cone.
Fig.\ref{FigRind} shows such a  situation.
It is interesting that the condition for residues has a 
solution on an intersection of the light-cone of the 
observer and the virtual path of a particle (dotted line in the left wedge). 
We skip the calculations and show the final results 
of the interference terms;
\begin{align}
&\langle \phi_{I}(x) \phi_h(y) \rangle
  +\langle \phi_h(x) \phi_{I}(y) \rangle \\ &= 
 \frac{ -i a e^2 x^i y^i }{ (4 \pi)^2 \rho_0(x)^2 \rho_0(y)^2 }
  \int \frac{d\omega}{2\pi} \frac{1}{1 -e^{-2\pi \omega/a}} 
  \label{EqInterfereScl} \notag \\
& \times \Bigl[
   e^{-i\omega (\tau_-^x -\tau_-^y)} h(-\omega) 
	\bigl( \frac{aL_x^2}{2 \rho_0(x)} -\frac{i\omega}{a} \bigr) \nonumber \\
& \qquad e^{-i\omega (\tau_-^x -\tau_-^y)}
	 - h( \omega) \bigl( \frac{aL_y^2}{2 \rho_0(y)} +\frac{i\omega}{a} \bigr) 
	 \nonumber \\  
& \quad + e^{-i\omega (\tau_+^x -\tau_-^y)} h(-\omega) 
	  \bigl( - \frac{a L^2_x}{2 \rho_0(x)} -\frac{i\omega}{a} \bigr)
	  Z_x (-\omega) \nonumber \\
& \quad - e^{-i\omega (\tau_-^x -\tau_+^y)} h(\omega) 
	  \bigl( - \frac{a L^2_y}{2 \rho_0(y)} +\frac{i\omega}{a} \bigr) 
	  Z_y (-\omega)
  \Bigr] \nonumber 
\end{align}
where
\begin{align}
& Z_x(\omega) = e^{\pi \omega /a} \theta (x^0-x^1) + \theta (x^1-x^0) \\
& L_x^2 = -x^\mu x_\mu + \frac{1}{a^2}, \quad 
  L_y^2 = -y^\mu y_\mu + \frac{1}{a^2}.
\end{align} 
\subsection{Partial Cancellation}
The correlation function of the inhomogeneous terms (\ref{EqIninScl}) 
depends only on $\tau_-$. 
The interference terms contain both of terms depending on $\tau_-$
and $\tau_+$; the first term in the parenthesis of (\ref{EqInterfereScl}) 
 depends only  on $\tau_-$, so it is the term that may cancel 
 the inhomogeneous terms (i.e. the Unruh radiation).
Using the relation
\begin{align}
  h(\omega) + h(-\omega) 
  &=
  \frac{e^2}{6\pi} (\omega^2 +a^2) |h(\omega)|^2,
\end{align}
one can show that a part of the interference terms
\begin{align}
& \frac{ i a e^2 x^i y^i }{ (4 \pi)^2 \rho_0(x)^2 \rho_0(y)^2 }
 \nonumber \\ 
&\times \int \frac{d\omega}{2\pi} 
        \frac{e^{-i\omega (\tau_-^x -\tau_-^y)}}{1 -e^{-2\pi \omega/a}} 
 \bigl(
   h(-\omega) \frac{i\omega}{a} + h( \omega) \frac{i\omega}{a} 
 \bigr)
 \end{align}
 cancels the first correction term of the inhomogeneous part in (\ref{EqIninScl}).
This term was obtained by taking a derivative of $e^{i \omega \tau_-^x}$ in $P(x,\omega)$. But note that the cancellation occurs only partially.
Furthermore, the $\tau_+$-dependent terms in the interference terms
cannot be canceled with the Unruh radiation. 
\subsection{The Energy Momentum Tensor}
Given the 2-point function, we can calculate 
the energy momentum tensor of the radiation 
\begin{align}
  \langle T_{\mu\nu} (x) \rangle 
  = 
 \langle :\partial_{\mu}\phi \partial_{\nu}\phi 
  -\frac{1}{2} g_{\mu\nu} \partial^{\alpha}\phi \partial_{\alpha}\phi: \rangle.
\end{align}
It is a sum of the classical and the fluctuation parts;
$
T_{\mu\nu} = T_{cl,\mu\nu} + T_{fluc,\mu\nu}.
$
The classical part is given by
\begin{align}
  T_{cl, \mu\nu}  
&\sim
  \frac{e^2 \partial_\mu \rho_0 \partial_\nu \rho_0 }
  {(4\pi)^2 \rho^4_0}.
\end{align}
It corresponds to the energy momentum tensor of the Larmor radiation.
 From (\ref{rhodef}) it can be seen to be  proportional to $a^2/r^2$ where
 $a$ is the acceleration and $r$ is the spacial distance from the particle
 to the observer.
$T_{fluc,\mu\nu}$ is the energy momentum of the additional radiation
\begin{align}
  \label{Tfluc}
&  T_{fluc,\mu\nu} 
  =
  \frac{ (x^i)^2 }{ \rho^2_0 } 
  \biggl[
    \Bigl(
      \frac{e^2}{\pi} I_m -  \frac{6ma^2 I_1 L_x^2}{\rho_0} 
     \Bigr) 
    T_{cl,\mu\nu}  
\nonumber  \\ &
	 	 -\frac{e^2 a^2 L_x^2}{(4\pi)^2 \rho_0^3} \Bigl(
	   m I_3 \ \partial_{\mu} \tau_-^x \partial_{\nu} \tau_-^x 
		 + \frac{2m I_1}{\rho_0 L_x^2} 
		  ( x_{\mu} \partial_{\nu} \rho_0 +x_{\nu} \partial_{\mu} \rho_0 ) 
\nonumber \\ &		  
  +\frac{e^2 I_m}{12\pi L_x^2} 
		  ( x_{\mu} \partial_{\nu} \tau_-^x +x_{\nu} \partial_{\mu} \tau_-^x )
 \nonumber \\ &
	  -\frac{e^2 I_m}{24\pi \rho_0} 
	   ( \partial_{\mu} \tau_-^x \partial_{\nu}\rho_0 
	   + \partial_{\nu} \tau_-^x \partial_{\mu}\rho_0 )
	 \Bigr)
  \biggr] 
\end{align}
where
$
 I_1 = \frac{3}{2mae^2}, 
 I_3  \sim \Omega_-^2 I_1 \ll a^2 I_1, 
 I_m = I_3+ a^2 I_1 \sim a^2 I_1.  
$
Hence, these terms originating from the fluctuating motion of the particle
is proportional to $a^3$, and
smaller by a factor of $a$ compared to the above Larmor radiation.
Though they have different angular distribution, there is an overall factor $(x_i^2)$ in front and
they vanish at the forward direction.  
Together with the long relaxation time discussed in \cite{IYZ2}, the detection of the Unruh radiation 
seems to be very difficult experimentally.



\begin{thebibliography}{9}   

\bibitem{IYZ} 
 S.~Iso, Y.~Yamamoto and S.~Zhang,
  arXiv:1011.4191 [hep-th].
  
\bibitem{IYZ2}
 S.~Iso, Y.~Yamamoto and S.~Zhang, in the same proceeding,
 ``Can we detect "Unruh radiation" in the high intensity laser?''
\bibitem{ChenTajima}
  P.~Chen and T.~Tajima,
  Phys.\ Rev.\ Lett.\  {\bf 83} (1999) 256.
 

%
%

\bibitem{JohnsonHu}  
  P.~R.~Johnson and B.~L.~Hu,
  arXiv:quant-ph/0012137.
  Phys.\ Rev.\  D {\bf 65} (2002) 065015
  [arXiv:quant-ph/0101001].
 P.~R.~Johnson and B.~L.~Hu,
  Found.\ Phys.\  {\bf 35}, 1117 (2005)
  [arXiv:gr-qc/0501029].

\bibitem{Sciama}
  D.~J.~Raine, D.~W.~Sciama and P.~G.~Grove,
  Proc.\ R.\ Soc.\ Lond.\ A (1991) 435, 205-215



\end{thebibliography}
\end{document}